\documentclass[format=sigconf,anonymous=false,review=false]{acmart}

\AtBeginDocument{%
  \providecommand\BibTeX{{%
    \normalfont B\kern-0.5em{\scshape i\kern-0.25em b}\kern-0.8em\TeX}}}

\copyrightyear{2022}
\acmYear{2022}
\setcopyright{acmcopyright}
\acmConference[SIGIR '22] {Proceedings of the 45th International ACM SIGIR Conference on Research and Development in Information Retrieval}{July 11--15, 2022}{Madrid, Spain.}
\acmBooktitle{Proceedings of the 45th International ACM SIGIR Conference on Research and Development in Information Retrieval (SIGIR '22), July 11--15, 2022, Madrid, Spain}
\acmPrice{15.00}
\acmISBN{978-1-4503-8732-3/22/07}
\acmDOI{10.1145/3477495.3531909}

\makeatletter
\def\blfootnote{\gdef\@thefnmark{}\@footnotetext}
\makeatother

\usepackage{graphicx}
\usepackage[caption=false]{subfig}
\usepackage{color}
\usepackage{algorithm}
\usepackage{algorithmic}

\usepackage{microtype}
\usepackage{amsmath,amsfonts}
\usepackage{mathtools}
\usepackage{mathrsfs}
\usepackage{bm}
\usepackage{booktabs, multirow}
\usepackage{makecell}
\usepackage{stfloats}
\usepackage{xcolor}
\usepackage{arydshln}
\usepackage{pifont}

\begin{document}
\fancyhead{}


\title{Towards Personalized Bundle Creative Generation with Contrastive Non-Autoregressive Decoding}

\author{Penghui Wei}
\affiliation{%
  \institution{Alibaba Group}
  \city{Beijing} 
  \country{China} 
}
\email{wph242967@alibaba-inc.com}

\author{Shaoguo Liu}
\affiliation{%
  \institution{Alibaba Group}
  \city{Beijing} 
  \country{China} 
}
\email{shaoguo.lsg@alibaba-inc.com}

\author{Xuanhua Yang}
\affiliation{%
  \institution{Alibaba Group}
  \city{Beijing} 
  \country{China} 
}
\email{xuanhua.yxh@alibaba-inc.com}

\author{Liang Wang}
\affiliation{%
  \institution{Alibaba Group}
  \city{Beijing} 
  \country{China} 
}
\email{liangbo.wl@alibaba-inc.com}

\author{Bo Zheng}
\affiliation{%
  \institution{Alibaba Group}
  \city{Beijing} 
  \country{China} 
}
\email{bozheng@alibaba-inc.com}

\begin{abstract}
Current bundle generation studies focus on generating a combination of items to improve user experience. In real-world applications, there is also a great need to produce bundle creatives that consist of mixture types of objects (e.g., items, slogans and templates) for achieving better promotion effect. We study a new problem named bundle creative generation: for given users, the goal is to generate personalized bundle creatives that the users will be interested in. To take both  quality and efficiency into account, we propose a contrastive non-autoregressive model that captures user preferences with ingenious decoding objective. Experiments on large-scale real-world datasets verify that our proposed model shows significant advantages in terms of creative quality and generation speed. 
\blfootnote{Corresponding author: S. Liu. }
\end{abstract}

%
%



\maketitle

\section{Introduction}
Many online recommendation systems begin to recommend the combination of multiple items as a bundle to users. E-commerce sites promote personalized bundled items to enhance user engagement~\cite{zhu2014bundle,bai2019personalized,gong2019exact}. The online gaming platform Steam offers sets of games to players with a discounted rate~\cite{pathak2017generating}. 
Recommending bundles other than individual items brings benefits: for sellers, the items obtain more exposure opportunities when they are displayed with other items; for customers, their interests are broadened and satisfied better. Therefore, bundle recommendation as well as generation has attracted increasing research attention in recent years~\cite{vijaikumar2020gram,chang2021bundle}. 

\begin{figure}[t]
\centering
\centerline{\includegraphics[width=1.0\columnwidth]{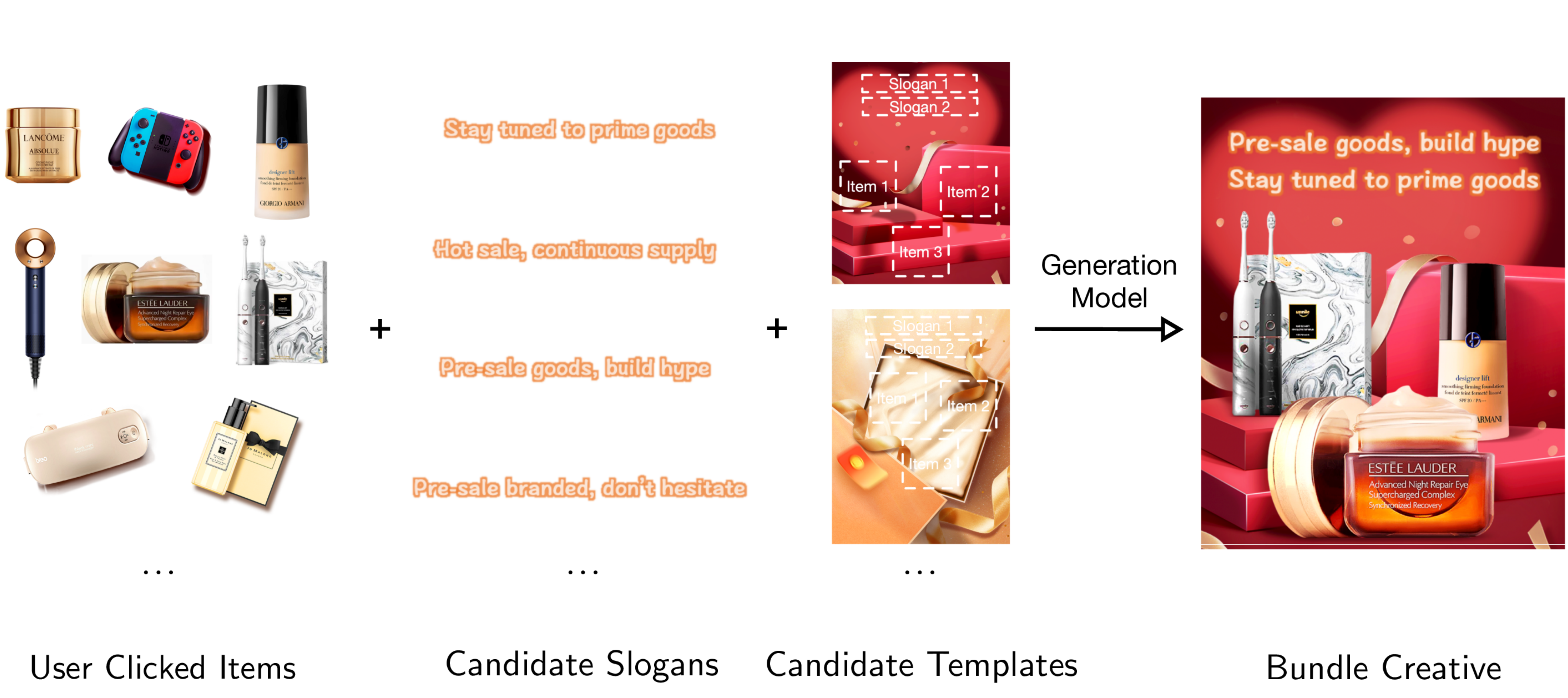}}
\vspace{-0.6em}
\caption{An illustration of bundle creative. Unlike ordinary bundles that contain items, a bundle creative consists of mixture types of objects: items, slogans and a template.}
\label{fig1:bundle}
\end{figure}

To achieve better promotion effect, there is also a great need to produce bundle creatives that consist of mixture types of objects (e.g., for online advertising and visual recommendation~\cite{deldjooleveraging}). 
Figure~\ref{fig1:bundle} shows an example that consists of three types of objects: a template (containing three item slots and two slogan slots), three items and two slogans. 
For each bundle, its creative is the content carrier that will be displayed to users. Appropriate bundle creatives can attract customers  and bring improvement on user experience. 
Thus sellers will benefit from the service of automatically generating bundle creatives provided by platforms, especially for small and medium-sized sellers which may not have much experience.

This paper studies a new problem named personalized bundle creative generation: for given users as well as their context information (e.g., historical interactions), the goal is to generate personalized bundle creatives that the users will be interested in. 
This problem is faced with the following challenges: 
(i) {Creative-level quality}: We not only need to mine the combination of items from users' historical interactions, but also consider the  creative as a whole to meet user preference and ensure creative quality. 
(ii) {Generation efficiency}: The generation speed of bundle creatives is also a key factor of models, especially in real-time service. Current bundle generation models usually employ sequential mechanism to produce each item step-by-step~\cite{bai2019personalized,gong2019exact,chang2021bundle}, which is inefficient to real-time applications and the quality is affected by item order. 
Such challenges make it difficult for existing bundle generation models (i.e., taking item as the only type of object) to tackle the bundle creative generation problem. 
Although there are a few approaches for ad creative optimization~\cite{chen2021automated}, they aim to composite multiple elements of a creative for {a given item} (that is, this task does not need to mine item or item combination from user context). 

With the aim of taking both creative-level quality and generation efficiency into account, in this paper we propose a \underline{con}trastive \underline{n}on-\underline{a}utoregressive decoding model (\textsf{Conna}) 
for bundle creative generation, which captures user preferences  with ingenious decoding objective. 
Our \textsf{Conna} model is an encoder-decoder architecture. 
A type-aware encoder adopts self-attention to learn the representations for mixture types of candidate objects.
A non-autoregressive decoder generates all objects of a creative in parallel, which is unaware of object ordering and furthest improves decoding efficiency. To ensure creative-level quality that meets user preferences, the  \textsf{Conna} model is optimized via a contrastive learning objective, which measures the quality of generated creatives by considering user positive/negative feedbacks. The main contributions are:
\begin{itemize}
    \item To our knowledge, this work is the first attempt that studies the problem of bundle creative generation, which makes effort to improve promotion effect in real world applications. We will release our datasets to facilitate further research. 
    \item We propose a contrastive non-autoregressive decoding model (\textsf{Conna}), which takes both creative-level quality and generation efficiency into account for bundle creative generation.
    \item  Experiments  verify that our \textsf{Conna} model shows significant advantages in terms of creative quality and generation speed. 
\end{itemize}

\begin{figure}[t]
\centering
\centerline{\includegraphics[width=\columnwidth]{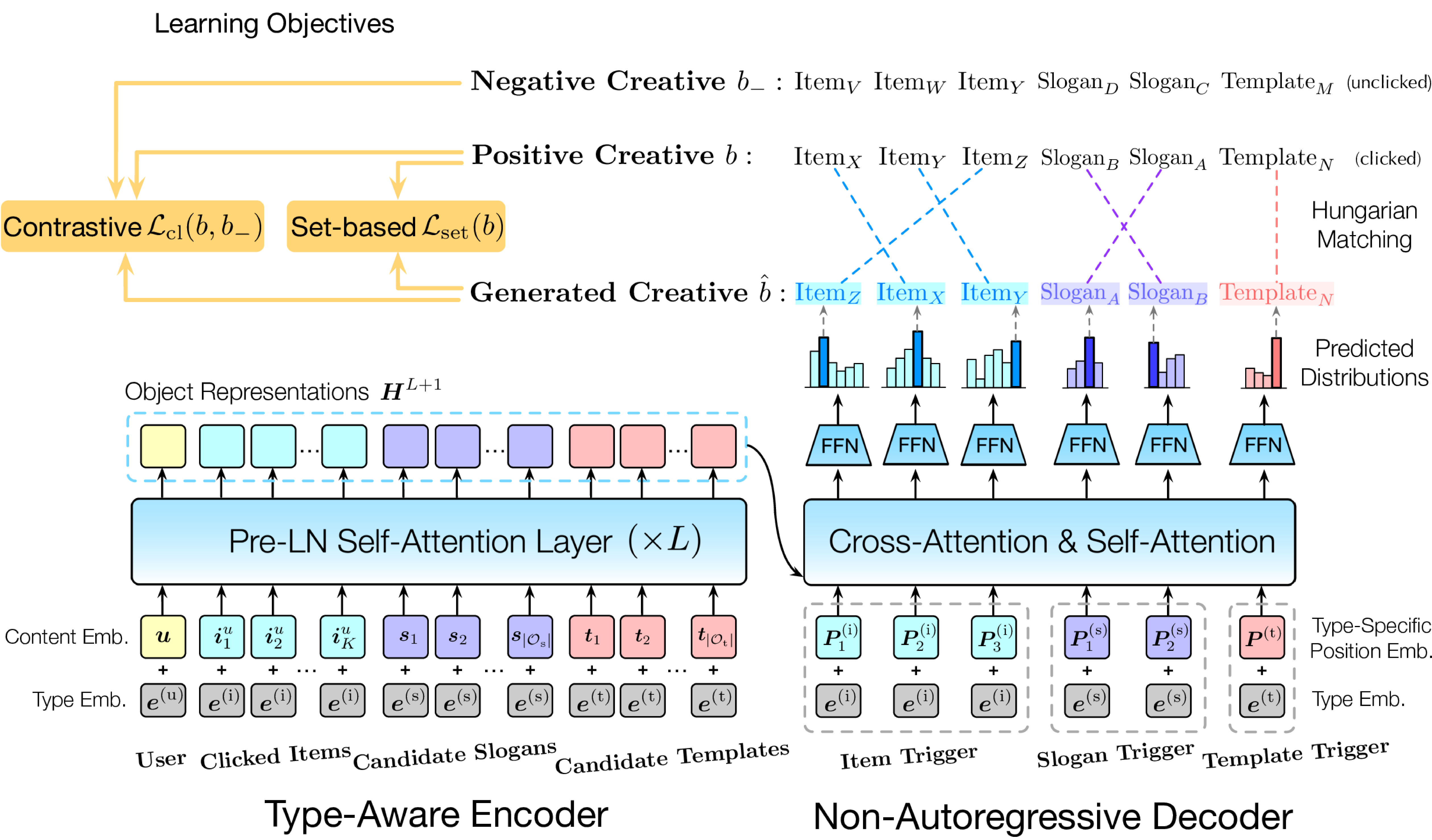}}
\vspace{-1em}
\caption{Overview of our contrastive non-autoregressive model \textsf{Conna} for personalized bundle creative generation.}
 \vspace{-1em}
\label{fig:overview}
\end{figure}

\section{Problem Definition}
Let $\mathcal O=\mathcal O_\mathrm{i}\cup \mathcal O_\mathrm{s}\cup \mathcal O_\mathrm{t}$ denote the candidate object set, where $\mathcal O_\mathrm{i}$ / $\mathcal O_\mathrm{s}$ / $\mathcal O_\mathrm{t}$ denotes candidate items / slogans / templates.\footnote{The problem can be generalized to any mixture types of objects.} 
For each bundle creative, we assume that the number of items / slogans / templates is $I$ / $S$ / $T$ (for the one in Figure~\ref{fig1:bundle}, $I=3, S=2$ and $T=1$). 

We have an interaction matrix between users and bundle creatives, where each element denotes whether the user  has positive feedback (i.e., click) to the bundle creative. The goal of personalized bundle creative generation is to learn a model $p(b\mid u, \mathcal{O})$: given user $u$'s context information and candidate objects $\mathcal{O}$, the model produces a bundle creative $b$ that the user is most satisfied with it. Consider that the size of candidate item set $\mathcal O_\mathrm{i}$ is usually large,  following~\cite{bai2019personalized} we resort to the user's historical clicked items $\mathcal O_{{\mathrm i}}^{u}$, other than the whole set $\mathcal O_\mathrm{i}$. Thus it is possible that a few items in $b$ are not contained in $\mathcal O^u_{\mathrm i}$.

\section{Proposed Model}
We propose a contrastive non-autoregressive model \textsf{Conna} for bundle creative generation. 
Figure~\ref{fig:overview} gives an overview of \textsf{Conna}, which contains a type-aware encoder and a non-autoregressive decoder to improve generation efficiency. It is optimized via a contrastive learning objective to ensure creative-level quality.

\subsection{Encoding Mixture Types of Candidates}
The input of \textsf{Conna} model contains four parts: the user $u$, the historical clicked item set $\mathcal O_{{\mathrm i}}^{u}=\left\{i^{u}_1,i^{u}_2,\ldots, i^{u}_K\right\}$ of this user, candidate slogan set $\mathcal O_\mathrm{s}=\left\{s_1,s_2,\ldots,s_{|\mathcal O_\mathrm{s}|}\right\}$ and candidate template set $\mathcal O_\mathrm{t}=\left\{t_1,t_2,\ldots,t_{|\mathcal O_\mathrm{t}|}\right\}$. 
For convenience, we denote the input as a unified sequence $\left(u;i^{u}_1,\ldots, i^{u}_K;   s_1,\ldots,s_{|\mathcal O_\mathrm{s}|};   t_1,\ldots,t_{|\mathcal O_\mathrm{t}|}  \right)$, next we will show that the encoder of \textsf{Conna} is unaware of such ordering of objects in a bundle creative. 

\textsf{Conna} employs a type-aware encoder to learn representations for mixture types of candidates, which consists of an embedding layer and several self-attention layers~\cite{vaswani2017attention}. 
Specifically, for each object of the input sequence, the embedding layer computes two embeddings to represent the object. 
The first is content embedding: for user, item, slogan and template IDs, we have four content embedding matrices $\bm U,\bm I,\bm S$ and $\bm T$ that transform each object ID to its embedding. 
The second is type embedding: we maintain a type embedding matrix $\bm E=\left[ \bm e^{(\mathrm{u})}; \bm e^{(\mathrm{i})}; \bm e^{(\mathrm{s})}; \bm e^{(\mathrm{t})}\right]\in \mathbb R^{d\times 4}$, where each embedding is for one type (i.e., user, item, slogan and template). For each object in the input sequence, the embedding layer adds its content embedding and type embedding:
\begin{equation}
\small
\begin{aligned}
\bm{\hat u}&= \bm u + \bm e^{(\mathrm{u})}\,,\qquad\qquad\qquad\quad 
\bm{\hat{i}}_k^{u}= \bm i^{u}_k + \bm e^{(\mathrm{i})}\ \ (k \in [1, K])\,,\\
\bm{\hat s}_i&= \bm s_i + \bm e^{(\mathrm{s})}\ \ (i \in [1, |\mathcal O_{\mathrm s}|])\,,\quad 
\bm{\hat t}_j= \bm t_j + \bm e^{(\mathrm{t})}\ \ (j \in [1, |\mathcal O_{\mathrm t}|])\,,
\end{aligned}
\end{equation}
where $\bm u,\bm i_k^{u},\bm s_i$ and $\bm t_j$ denote content embeddings for $u, i_k^u, s_i$ and $t_j$. 

The encoder then adopts $L$ self-attention layers to learn representations for each candidate object. Let $\bm H^l\in\mathbb R^{d\times (1+K+|\mathcal O_{\mathrm s}|+|\mathcal O_{\mathrm t}|)}$ denote the $l$-th layer's input, 
where the input of the first layer is $\bm H^1=\left(\bm{\hat u}, \bm{\hat{i}}_1^{u},\ldots,\bm{\hat{i}}_K^{u}, \bm{\hat s}_1,\ldots, \bm{\hat s}_{|\mathcal O_\mathrm{s}|}, \bm{\hat t}_1,\ldots, \bm{\hat t}_{|\mathcal O_\mathrm{t}|} \right)$. We use Pre-LN self-attention operation~\cite{klein2017opennmt} to produce output $\bm H^{l+1}$, which applies layer normalization before multi-head attention and position-wise feed-forward network to achieve faster convergence. 
The encoder of \textsf{Conna} produces representations $\bm H^{L+1}$ for each input object,\footnote{Although we may design a sophisticated encoder (e.g., based on graph neural networks~\cite{chang2021bundle,wei2021graph}), the main focus of this work is  the generation process, thus we employ a  simple-yet-effective design of encoder.} which is unaware of each object's position in the input sequence.

\clearpage
\subsection{Non-Autoregressive Creative Generation}
After encoding candidate objects, the decoder generates mixture types of objects forming a bundle creative to meet user preferences. 
Let $b=\left\{b^{(\mathrm i)}_1,b^{(\mathrm i)}_2,\ldots,b^{(\mathrm i)}_I; b^{(\mathrm s)}_1,b^{(\mathrm s)}_2,\ldots,b^{(\mathrm s)}_S; b^{(\mathrm t)}\right\}$ denote a bundle creative that the user $u$ has positive feedback, where $b^{(\mathrm i)}_*$, $b^{(\mathrm s)}_*$ and $b^{(\mathrm t)}$ denote item, slogan and template, respectively. A few items may not be contained in user historical clicked item set $\mathcal O^u_{\mathrm i}$. 

Traditional autoregressive decoder generates each object one-by-one, which factorizes the generation probability of $b$ to the multiplying of conditional probabilities. 
However, the factorization is affected by both the order of \{items, slogans, template\} and the order within items / slogans, yet we do not have a "ground-truth" ordering in fact. 
Besides, the efficiency is limited, because it generates each object conditioned on previously generated ones.

\subsubsection{\textbf{Non-Autoregressive Decoder Architecture}}
We furthest improve efficiency via modeling the bundle creative probability as: 
\begin{equation}
\small
    p(b\mid u, \mathcal O) = \prod_{j=1}^{I}p(b^{(\mathrm i)}_j\mid u, \mathcal O) \cdot \prod_{j=1}^{S}p(b^{(\mathrm s)}_j\mid u, \mathcal O) \cdot p(b^{(\mathrm t)}\mid u, \mathcal O) 
\end{equation}
Based on this factorization, the generation is unaware of the ordering, and at inference time it computes all distributions in parallel. 

Let $B=I+S+1$ denote object number. For convenience, we define that $I$ items are generated from the 1st to $I$-th positions, $S$ slogans are generated from the $(I+1)$-th to $(I+S)$-th positions, and the template is generated by the $B$-th position. 
In fact our decoder is not affected by this order due to non-autoregressive nature. 

Specifically, the decoder architecture of our \textsf{Conna} model consists of an embedding layer, several attention layers and an output layer. 
The embedding layer aims to produce $B$ ``trigger embeddings'' that guide the generation process. The trigger embedding of the $j$-th position ($j\in[1, B]$) is the sum of two parts: 
\begin{itemize}
    \item type embedding, which represents the object type of this position (e.g., for $1\leq j\leq I$ we take $\bm e^{(\mathrm{i})}$ from $\bm E$ because the object type is item; for $I+1\leq j\leq I+S$ we take $\bm e^{(\mathrm{s})}$).
    \item type-specific positional embedding, where we maintain three positional embedding matrices $\bm P^{(\mathrm i)}\in\mathbb R^{d\times I}, \bm P^{(\mathrm s)}\in\mathbb R^{d\times S}$ and $\bm P^{(\mathrm t)}\in\mathbb R^{d\times 1}$ for all types. For $1\leq j\leq I$, we take the $j$-th column from $\bm P^{(\mathrm i)}$ to represent that this position need to generate the $j$-th item. Similarly, for $I+1\leq j\leq I+S$, we take the $(j-I)$-th column from $\bm P^{(\mathrm s)}$ to represent that this position need to generate the $(j-I)$-th slogan.
\end{itemize}
The use of type-specific positional embedding ensures that the decoder can distinguish different objects from the same type during generation, avoiding the situation of generating repeat objects. 

Then attention layers employ self-attention and encoder-decoder cross attention to learn each position's representation~\cite{vaswani2017attention}. We remove causal mask in standard Transformer decoder, because we no longer need prevent to attend previously positions. In contrast, the decoder considers pair-wise relationships to globally modeling. 

Finally, the output layer employs position-wise feed-forward operation with softmax to compute each position's probability distribution over the candidate set of this position's object type. Let $\left\{\bm{\hat b}^{(\mathrm i)}_1, \ldots, \bm{\hat b}^{(\mathrm i)}_I; \bm{\hat b}^{(\mathrm s)}_1,\ldots,  \bm{\hat b}^{(\mathrm s)}_S; \bm{\hat b}^{(\mathrm t)}\right\}$ be each position's predicted distribution, where $\bm{\hat b}^{(\mathrm i)}_*\in\mathbb R^{|\mathcal O_{\mathrm i}|}$, $\bm{\hat b}^{(\mathrm s)}_*\in\mathbb R^{|\mathcal O_{\mathrm s}|}$ and $\bm{\hat b}^{(\mathrm t)}\in\mathbb R^{|\mathcal O_{\mathrm t}|}$.

\subsection{Optimization}
A straightforward objective is the independent cross-entropy (XEnt) losses of all positions by comparing predicted distributions and $b$: 
\begin{equation}\label{loss:seq}
\small
   \mathcal L^{(\mathrm i)}+\mathcal L^{(\mathrm s)}+\mathcal L^{(\mathrm t)}= \sum_{j=1}^I\mathsf{XEnt}\left(\bm{\hat b}^{(\mathrm i)}_j, b^{(\mathrm i)}_j\right)+\sum_{j=1}^{S}\mathsf{XEnt}\left(\bm{\hat b}^{(\mathrm s)}_j, b^{(\mathrm s)}_j\right)+\mathsf{XEnt}\left(\bm{\hat b}^{(\mathrm t)}, b^{(\mathrm t)}\right)
\end{equation}
However, in a bundle creative, because objects from same type are unordered, such optimization will penalize the true predictions that reorder items/slogans: for instance, if we have:
\vspace{-0.5em}
\begin{equation*}
\small{
    b^{(\mathrm i)}_1=\mathrm{item}_X,\ \  b^{(\mathrm i)}_2=\mathrm{item}_Y,\ \  b^{(\mathrm i)}_3=\mathrm{item}_Z
}
\vspace{-1em}
\end{equation*}
\begin{equation*}
\small{
    \arg\max\bm{\hat b}^{(\mathrm i)}_1=\mathrm{item}_Z, \ \arg\max\bm{\hat b}^{(\mathrm i)}_2=\mathrm{item}_X, \ \arg\max\bm{\hat b}^{(\mathrm i)}_3=\mathrm{item}_Y
}
\end{equation*}
the model produces true prediction, but the above learning objective will suggest that all positions are wrongly predicted.

\subsubsection{\textbf{Set-based Learning Objective}}
To avoid above inaccurate penalty, inspired by~\cite{carion2020end,du2021order} we propose a set-based learning objective for training the \textsf{Conna} model.  
Formally, consider the item type, we construct all possible \textit{permutations} of $I$ items in $b$ as a permutation space $\mathbf{  B}^{(\mathrm i)}=\left\{\mathcal B^{(\mathrm i)}_1,\ldots,\mathcal B^{(\mathrm i)}_{I!}\right\}$, where the item set of each permutation $\mathcal B^{(\mathrm i)}_{*}$ is same (i.e., $\left\{b^{(\mathrm i)}_1, \ldots, b^{(\mathrm i)}_I\right\}$). 
We employ Hungarian algorithm~\cite{kuhn1955hungarian} to efficiently search a permutation from $\mathbf{  B}^{(\mathrm i)}$ that has minimal XEnt value: 
\begin{equation}
    \mathcal L^{(\mathrm i)}_{\mathrm{set}}(b) = \min\left\{ \mathcal L^{(\mathrm i)}\left(\mathcal B^{(\mathrm i)}_*\right)\right\}_{\mathcal B^{(\mathrm i)}_*\in {\mathbf B}^{(\mathrm i)}}
\end{equation}
where $\mathcal L^{(\mathrm i)}\left(\mathcal B^{(\mathrm i)}_*\right)$ means that we use the permutation $\mathcal B^{(\mathrm i)}_*$ instead of original $\left(b^{(\mathrm i)}_1,b^{(\mathrm i)}_2,\ldots,b^{(\mathrm i)}_I\right)$ as ground-truth to compute the $\mathcal L^{(\mathrm i)}$ in Equation~\ref{loss:seq}. For slogan type, $\mathcal L^{(\mathrm s)}_{\mathrm{set}}$ can be similarly defined.

Through the set-based objective for all types of objects:
\begin{equation}\label{loss:set}
    \mathcal L_{\mathrm{set}}(b) = \mathcal L^{(\mathrm i)}_{\mathrm{set}}(b)+\mathcal L^{(\mathrm s)}_{\mathrm{set}}(b)+\mathcal L^{(\mathrm t)}(b)    
\end{equation}
the optimization procedure does not penalize any true predictions.

\subsubsection{\textbf{Contrastive Learning Objective}}
With the aim of considering the generated creative as a whole to ensure creative-level quality, we incorporate the bundle creatives that are exposed to the user $u$ but are not clicked as negative instances $\{b_{-}\}$. Specifically, we propose to explicitly model the generation probabilities of both positive and negative bundle creatives during optimization. 
A margin-based loss is used to maximize the probability gap of generating positive bundle creative $b$ and negative ones $\{b_{-}\}$:
\begin{equation}
    \mathcal L_{\mathrm{cl}}(b, b_{-}) = \sum_{b_{-}} \max\biggl\{ 0, - \biggl( \mathcal L_{\mathrm{set}}(b_{-}) -  \mathcal L_{\mathrm{set}}(b)  \biggr) + \gamma \biggr\}
\end{equation}
where $\gamma$ denotes the margin hyperparameter. 

Overall, the \textsf{Conna} model is optimized via the following objective, where $\lambda$ balances two terms:
\begin{equation}
    \mathcal L_{\mathrm{set}}(b) + \lambda \mathcal L_{\mathrm{cl}}(b, b_{-})\,.
\end{equation}

\subsubsection{\textbf{Inference}}
At inference time, the \textsc{Conna} model produces  bundle creative via generating all objects in parallel, where $\mathrm{argmax}$ operation is applied to each position's predicted distribution ($\bm{\hat b}^{(\mathrm i)}_*$, $\bm{\hat b}^{(\mathrm s)}_*$ and $\bm{\hat b}^{(\mathrm t)}$).

\section{Experiments}
\subsection{Experimental Setup}
\textbf{Datasets}\quad 
We collected real world user feedbacks from a large e-commerce advertising platform to construct large-scale bundle creative (BC) datasets (Figure~\ref{fig1:bundle} shows the style of BCs). 
The overall feedback log contains 7 million instances, where 100k of them are positive instances (i.e., clicked by users, in which each user has clicked on 1.03 BC on average).  
We split it to training/development/test sets via timestamp with the proportion of 7:1:2. 
To investigate the dataset size's impact, we perform sampling to training set with a ratio of 30\% to obtain a small version, named BC-S dataset, and the full dataset is named BC-L.

\textbf{Competitors}\quad 
We compare the following models designed for bundle creative generation. They use same encoder architecture, and the differences are decoder and learning objective. 

(i) \textit{\textsf{MultiDec}}: It employs three independent autoregressive decoders (RNNs) to produce items, slogans and template respectively. 

(ii) \textit{\textsf{UnifiedDec}}: It employs a unified decoder to autoregressively generate items, then slogans, and finally template.\footnote{We empirically find that this ordering performs best. See section~\ref{experiments:ordering}.} At each time step we need to generate a specific type of object, thus we only generate from this type's vocabulary and mask other types' vocabularies to avoid generating wrong type. 

(iii) \textit{\textsf{PointerNet}}: Similar to the approach in~\cite{gong2019exact}, a pointer mechanism~\cite{vinyals2015pointer} is equipped to the decoder in \textit{\textsf{UnifiedDec}}, which can copy element from input side by a gating mechanism. 

(iv) \textit{\textsf{RL-PointerNet}}: It follows~\cite{gong2019exact} that trains a reward estimator to represent creative-level quality, and fine-tunes the \textsf{PointerNet} via REINFORCE~\cite{williams1992simple}. 

(v) \textit{\textsf{RL-Transformer}}: It replaces the decoder of \textit{\textsf{RL-PointerNet}} from RNN to standard Transformer decoder. 

(vi) \textit{\textsf{Conna}} is our proposed non-autoregressive model. 

For fair comparison, all models have same configuration (layer number $L=3$, dimension $d=256$, the size of user historical clicked items is 50). For each positive instance we sample up to three negative ones for $\mathcal L_{\mathrm{cl}}$, and we set $\gamma=1$ and $\lambda=0.5$.

\textbf{Evaluation Metrics}\quad 
Let $b$ ($\hat b$) denote gold (generated) bundle creative, where $b^{(\mathrm i)}$/$b^{(\mathrm s)}$/$b^{(\mathrm t)}$ denotes the item set / slogan set / template of $b$, and $\hat b^{(\mathrm i)}$/$\hat b^{(\mathrm s)}$/$\hat b^{(\mathrm t)}$ denotes that of $\hat b$. We evaluate each model from three aspects: \textit{quality}, \textit{diversity} and \textit{efficiency}. 

(1) For evaluating the creative quality, we design HitRatio-based metric. Specifically, a weighted sum formulation is defined as:
\begin{equation*}
    \mathrm{HitRatio}=\frac{I}{I+S+1}\cdot\frac{|b^{(\mathrm i)} \cap \hat b^{(\mathrm i)} |}{I} + \frac{S}{I+S+1}\cdot\frac{|b^{(\mathrm s)} \cap \hat b^{(\mathrm s)} |}{S} + \frac{|b^{(\mathrm t)} \cap \hat b^{(\mathrm t)} |}{I+S+1}
\end{equation*}

(2) To evaluate item diversity in a generated creative, we design the metric $\mathrm{Diversity}$:
\begin{equation*}
\mathrm{Diversity}=\frac{1}{|I|(|I|-1)} \sum_{ \hat b^{(\mathrm i)}_j, \hat b^{(\mathrm i)}_{j'} \in \hat b^{(\mathrm i)} }\left(1-\mathbb{I}\left( \hat b^{(\mathrm i)}_{j} = \hat b^{(\mathrm i)}_{j'} \right)\right) 
\end{equation*}

This metric shows the basic utility the generative bundle creative, because repeated object is unsatisfactory. 

(3) To evaluate efficiency of a model, we compare the speedup over the autoregressive model \textsf{RL-Transformer} when decoding a single bundle creative.

\begin{table}[t]
\small
\caption{Results on personalized bundle creative generation (HR. and Div. denote HitRatio and Diversity, respectively). The improvement of \textsf{Conna} on HR. is statistically significant to the second-best model.}
\centering
\begin{tabular}{lccccc}
\toprule
\multirow{2}*{\textbf{Model}} & \multicolumn{2}{c}{\textbf{BC-S Dataset} } & \multicolumn{2}{c}{\textbf{BC-L Dataset} }  & \\
\cmidrule(lr){2-3}\cmidrule(lr){4-5}
(Same Encoder)   & HR.  & Div.  &  HR. & Div. & Speedup \\
\midrule
\textsf{MultiDec}     &   0.3806   &  0.9147 &  0.4576 & 0.9349  &  1.6$\times$ \\ 
\textsf{UnifiedDec}   &   0.4622  &  0.9483  &  0.5721 & 0.9596  & 1.3$\times$ \\ 
\textsf{PointerNet}   &   0.4776  &  0.9335  &  0.6079 & 0.9709  & 1.2$\times$ \\
\textsf{RL-PointerNet} &  0.5107  &  0.9474  &  0.6209 & 0.9726  & 1.2$\times$\\
\textsf{RL-Transformer}  & 0.5317  & 0.9542  &  0.6411 & \textbf{0.9825}  &  1$\times$  \\
\cmidrule(lr){1-6}
\textsf{Conna} (Ours)  & \textbf{0.5431} &\textbf{ 0.9642}   & \textbf{0.6564} & 0.9802 & \textbf{2.8$\times$} \\ 
\bottomrule
\end{tabular}
\label{results:main}
\end{table}

\subsection{Main Results}
Table~\ref{results:main} shows the comparison results of all comparative models for personalized bundle creative generation. By comparing \textsf{MultiDec} and \textsf{UnifiedDec}, the results show that independent decoders cannot capture the relationship among object types during generation, and \textsf{MultiDec} performs poorly on creative quality. Thus generating mixture types of objects is a challenging problem. 
\textsf{PointerNet} employs pointer to select elements from encoder side during decoding, and outperforms \textsf{UnifiedDec}. This verifies that the generation performance benefits from the improvement of decoding mechanism, as shown in previous work~\cite{gong2019exact}. 

By training an extra reward estimator that provides creative-level quality, \textsf{RL-PointerNet} further outperforms \textsf{PointerNet} by a large margin through a fine-tuning process with the objective of maximizing expected reward of the generated bundle creatives. This shows that taking creative-level quality into account for creative generation is effective to improve the overall performance of bundle creative generation. 
\textsf{RL-Transformer} employs a self-attention based decoder to model the dependency among generated objects, and performs much better than traditional RNN-based \textsf{RL-PointerNet}. This demonstrates that the Transformer architecture is suitable to bundle creative generation task which needs to decode mixture of types of objects.

Our proposed \textsf{Conna} model employs a non-autoregressive decoder that furthest improves decoding efficiency for bundle creatives, and considers creative-level quality during training with a contrastive learning objective. It achieves the best performance among all competitors in terms of both generation quality (HitRatio) and effiency (Speedup). This indicates that the \textsf{Conna} model possesses the ability of balancing effectiveness and efficiency for bundle creative generation task. 

Besides, note that our \textsf{Conna} model also has advantage in terms of training efficiency compared to RL-based models. The contrastive learning objective is coupled with the set-based generation objective in an end-to-end manner, which is unlike other RL-based comparative models that need three steps to produce the final model: 1) train a reward estimator, 2) train a generation model, 3) finally fine-tune the trained model guided by reward estimator. Therefore, the proposed \textsf{Conna} model shows advantages on multiple aspects compared to other models.

\subsection{Discussion}
\subsubsection{\textbf{Ablation Study}}
The key of \textsf{Conna} is two-fold: the first is the contrastive objective that utilizes negative bundle creatives during optimization, and the second is the set-based objective that does not penalize any true predictions during optimization. 
To verify the effectiveness of them in our \textsf{Conna}, we conduct ablation experiments, in which the first variant removes the constrastive objective during training, and the second variant further replaces the set-based objective to the independent XEnt objective of Equation~\ref{loss:seq}.

Table~\ref{results:ablation} shows the results.
We observe that the contrastive objective contributes to the overall quality of \textsf{Conna}. Moreover, by using the set-based objective, the generation performance achieves further improvement. In terms of the input design of our decoder, type-specific positional embedding also brings uplift to the HitRatio, demonstrating its effect to decoder input.

\begin{table}[t]
\caption{Ablation experiments of the \textsf{Conna} model.}
\small
\centering
\begin{tabular}{lcc}
\toprule
\textbf{Model} & HR.  & Div.\\
\midrule
\textsf{Conna} (full model)                   & 0.6564 & 0.9802 \\ 
\hdashline[2pt/1.2pt] 
\quad w/o contrastive objective &  0.6194 & 0.9822\\
\quad w/o set-based objective, w/ independent XEnt   &  0.5893 & 0.9822 \\
\bottomrule
\end{tabular}
\label{results:ablation}
\end{table}

\begin{figure}[t]
\centering
\centerline{\includegraphics[width=0.9\columnwidth]{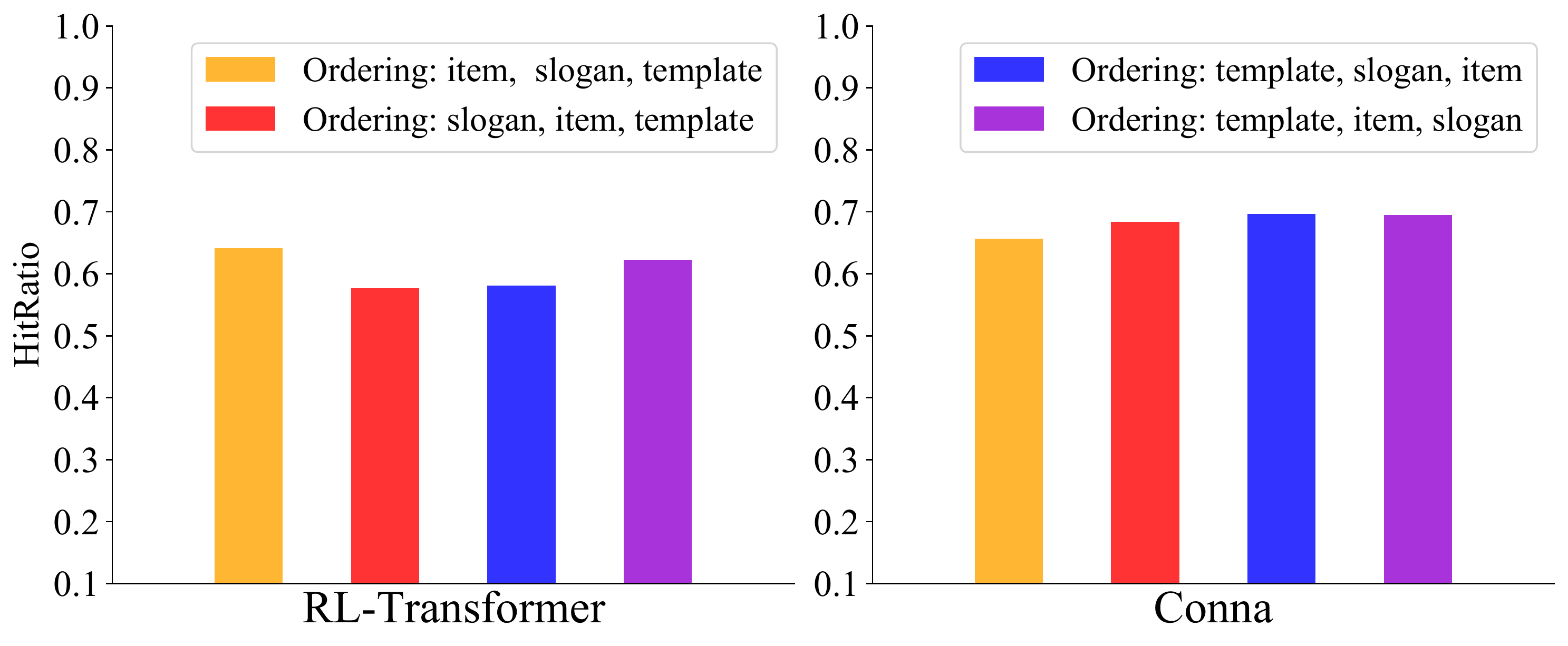}}
\caption{Results with different ordering of object types.}
\label{fig3:bundle}
\end{figure}

\subsubsection{\textbf{Analysis on Ordering of Object Types}}\label{experiments:ordering}

Autoregressive-based models need a pre-defined ordering of object types to produce each object in a bundle creative one-by-one. 
In contrast, our \textsf{Conna} is unaware of that due to the non-autoregressive nature. 

We compare the performance of two competitors w.r.t. different ordering of object types. As the results in Figure~\ref{fig3:bundle}, we observe that \textsf{Conna} is indeed not affected by the ordering, verifying the advantage of non-autoregressive decoding for bundle creative generation.

\section{Conclusion}
This work studies a new problem of personalized bundle creative generation. 
To take both quality and efficiency into account, we propose a contrastive non-autoregressive model that captures user preferences with ingenious decoding objective. 
Results verify that it shows significant advantages in terms of creative quality and generation speed. 

In future work, we shall explore how to avoid conflicting and unfavorable items to be grouped together during generation. 

\section*{Acknowledgments}
We thank all the anonymous reviewers to their insightful comments.

\bibliographystyle{ACM-Reference-Format}
\bibliography{src-shortversion.bib}

\end{document}